\documentstyle[12pt,epsfig,axodraw]{article}
\renewcommand\caption{}

\title{Dynamical Symmetry Breaking in Spaces with Constant Negative Curvature
}

\author{{\sl E. V. Gorbar} \\
{\sl{Instituto de Fisica Teorica, 01405-900 Sao Paulo, Brazil} \dag}
}

\date{}

\begin{document}

\maketitle

\vfill

\begin{abstract} 

By using the Nambu-Jona--Lasinio model, we study dynamical symmetry breaking in spaces with constant
negative curvature.
We show that the physical reason for zero value of critical coupling value $g_c = 0$ in these spaces
is connected with the effective reduction of dimension of spacetime
$1 + D \to 1 + 1$
in the infrared region, which takes place
for any dimension $1 + D$. Since the Laplace--Beltrami operator has a gap in spaces with constant negative curvature,
such an effective reduction for scalar fields is absent and there are not problems with radiative corrections due
to scalar fields.
Therefore, dynamical symmetry breaking with the effective
reduction of the dimension of spacetime for fermions in the infrared region is consistent with the
Mermin--Wagner--Coleman theorem, which forbids spontaneous symmetry breaking in (1 + 1)-dimensional spacetime.
 
\end{abstract}
\dag On leave of absence from Bogolyubov Institute for Theoretical Physics, 252143,
Kiev, Ukraine

\vspace{2mm}

\vfill
\eject

\newpage

\newpage
\section{Introduction.}

\vspace{1cm}

It is well known that dynamical symmetry breaking (DSB) presents an attractive alternative to the Higgs mechanism
of electroweak symmetry breaking in Standard Model and in a natural way solves the so called hierarchy problem
connected with
the quadratic divergence of the mass of the Higgs boson (for a general introduction to DSB see \cite{book}).
Morever, DSB allows at least in principle
to deduce all relevant parameters of symmetry breaking.  However, usually DSB requires strong coupling
($\alpha_c \ge 1$)
that essentially restricts the choice of models that can be used and it makes also
quantative studying of DSB a difficult problem. Physically, it is easy to understand why
$\alpha_c \ge 1$. For the state with the condensate of fermion-antifermion pairs to have lower energy than
the trivial
vacuum, it is necessary that energy of the corresponding
fermion-antifermion bound state be negative. Then, for example, in QED, in view of the uncertainty principle,
it imples $\alpha_c \ge 1$. Therefore, it is very interesting to consider situations when DSB takes place
in the regime of weak coupling ($\alpha_c \approx 0$).

Two examples of DSB in the regime of weak coupling are known. The first is symmetry breaking in the presence
of the Fermi surface (i.e. chemical potential is nonzero).
In this case, as well known from the Bardeen--Cooper--Schrieffer theory of superconductivity  
\cite{BCS}, a bound state forms for any (however small) attraction between fermions. The effective field theory
description of this phenomenon
based on the renormalization group was developed in \cite{Pol}, where it was shown that renormalization group
scaling takes place only in the direction perpendicular to the Fermi surface, therefore, from the viewpoint of
renormalization
group scaling the effective dimension of spacetime is 1 + 1. Since in two-dimensional
spacetime a bound state forms even in the case of arbitrary small attraction, we obtain that $\alpha_c = 0$ in this case.
(Note that this is one of the key ideas of QCD color superconductivity at finite density in the regime of weak coupling,
which has been actively studied in recent years \cite{Alf, Rap}).
The other example of DSB in the regime of weak coupling is DSB in external constant magnetic field. This phenomenon was
discovered in \cite{GMSh1,GMSh2}, where it was shown that chiral symmetry is dynamically broken
in the Nambu---Jona-Lasinio (NJL) model \cite{NJL} and QED
in external constant magnetic field $B$ for an arbitrary weak interaction,
i.e. the critical coupling constant is zero in
this case\footnote{The case of the discrete three-dimensional NJL model was considered in \cite{Kri} and the effect of
enhancement of the chiral condensate in supercritical ($g > g_c > 0$) phase of the four-dimensional NJL model
was studied in \cite{Kle}.}.
The essence of this strong magnetic catalysis \cite{GMSh1, GMSh2} is that electrons are
effectively (1 + 1)-dimensional when their energy is much less than the Landau gap $\sqrt{|eB|}$. The lowest Landau level
plays here the role similar to that of the Fermi surface in the BCS theory of superconductivity, leading to dimensional
reduction in dynamics of fermion pairing.

Recently another example of DSB in the regime of weak coupling was discovered. By using the NJL-type models it was
shown \cite{BK, IMO} that critical coupling constant is zero in spaces with constant negative curvature
(for an excellent review of DSB in curved
spacetime see \cite{Ina}), i.e. chiral symmetry is always broken for any $g > 0$ (note that the fact of impossibility
of keeping chiral invariance for free
massless fermions in spaces with negative constant curvature was also noted in \cite{Wil}).
The physical explanation of this very interesting fact is lacking. The authors of these works
calculated the effective potential for an order parameter and then showed that it has a nontrivial minimum for any $g > 0$.
To find an explanation of this result in more physical terms was one of the main motivations of the present work.
   
\vspace{1cm}

\section{The model.}

\vspace{1cm}

For our aims it is enough to consider the NJL model
in curved spacetime

\begin{equation}
     S = \int\! \sqrt{-g} d^{4}x \left[\sum_{k=1}^{N}
     \bar{\psi_k}i\gamma^{\mu}\nabla_{\mu}\psi_k
     + \frac{G}{2N} \left( (\sum_{k=1}^{N}\bar{\psi_k}\psi_k )^{2} -
     (\sum_{k=1}^{N}\bar{\psi_k}\gamma_5 \psi_k)^{2} \right)
     \right],
\end{equation}
where N is the number of flavors,
$g = \mbox{det} (g_{\mu\nu})$ the determinant of metric, $\nabla_{\mu} = \partial_{\mu} + i \omega^{ab}_{\mu}
\sigma_{ab}$ the covariant derivative with spin connection $\omega^{ab}_{\mu}$, and $\gamma^{\mu}$ matrices in curved
spacetime are expressed through the Dirac $\gamma^a$ matrices in flat spacetime with the help of vierbeins
$\gamma^{\mu} = V^{\mu}_{a}\gamma^{a}$.
The action (1) is invariant with respect to chiral transformations 
$\psi \rightarrow e^{i\gamma_5\beta}\psi$.
For practical calculations in four-fermion theories 
it is convenient to use the so called auxiliary 
field method \cite{Hub, Stra}, where
Lagrangian (1) is represented in the equivalent form
\begin{equation}
L= \sum_{k=1}^{N}\left(i\bar{\psi_k} \gamma^\mu \nabla_\mu\psi_k + \bar{\psi_k}(\sigma + i\gamma_5\pi)\psi_k\right)
- \frac{N}{2G}\sigma^2,
\end{equation}
where $\sigma$ and $\pi$ are auxiliary fields. If we integrate over $\sigma$ and $\pi$, we obtain the initial
action (1). If the field $\sigma$ acquires a nonzero vacuum expectation value, then obviously fermions acquire
mass and chiral symmetry is broken.
To find the effective action for the fields $\sigma$ and $\pi$, we integrate over the fermion fields. We obtain
(without loss of generality one can set $\pi = 0$ because it is always possible to restore the dependence on $\pi$
by requiring chiral symmetry of the effective action)
\begin{eqnarray}
\Gamma(\sigma_c) = -N\int \sqrt{-g}d^{4}x
                 \frac{\sigma_c^{2}}{2G}
                 -i\mbox{Ln Det}(i\gamma^{\mu}\nabla_{\mu}-\sigma_{c}),
\end{eqnarray}
where $\sigma_{c}(x)=<0|\sigma|0>$.
The effective potential $V(\sigma_c)$ (we set $\sigma_c(x)$ = const) is given by the expression
\begin{equation}
    V(\sigma_c)=-\frac{\Gamma(\sigma_c)}{\int\sqrt{-g}d^{4}x}.
\end{equation}
Furthermore,
\begin{eqnarray}
\mbox{Ln Det}(i\gamma^{\mu}\nabla_{\mu}-\sigma_c) = \mbox{TrLn}(i\gamma^{\mu}\nabla_{\mu}-\sigma_c) = \nonumber \\
\frac{1}{2} \mbox{TrLn}(i\gamma^{\mu}\nabla_{\mu}-\sigma_c) +
\frac{1}{2} \mbox{TrLn}(\gamma_5(i\gamma^{\mu}\nabla_{\mu}-\sigma_c)\gamma_5) = \nonumber \\
\frac{1}{2} \mbox{TrLn}\left((i\gamma^{\mu}\nabla_{\mu}-\sigma_c)(-i\gamma^{\nu}\nabla_{\nu}-\sigma_c) \right).
\end{eqnarray}
By using the Schwinger proper time method \cite{Sch}, we get the effective potential (we also perform the Wick rotation)
\begin{equation}
V(\sigma_c) = N\,\left(\frac{\sigma_c^{2}}{2G} + \frac{1}{2} \int_{\frac{1}{\Lambda^2}}^{\infty}\frac{ds}{s}\, \mbox{tr}
\,<x|e^{-sH}|x>\right),
\end{equation}
where $H = -(\gamma_E^{\mu}\nabla_{\mu})^2 + \sigma_c^2$ ($\gamma_E^{\mu}$ are Euclidean $\gamma$-matrices).
Consequently, the gap equation ($\frac{dV}{d\sigma_c}|_{\sigma_c=m} = 0$) is
\begin{equation}
1\,\, = \,\,G\int_{\frac{1}{\Lambda^2}}^{\infty}ds \, \mbox{tr} <x|e^{-sH}|x>.
\end{equation}
Thus, we need to find the diagonal heat kernel $\mbox{tr} <x|e^{-sH}|x>$ in spaces with
constant negative curvature. Before doing it we
first describe what these spaces are (for a very good introduction see \cite{Bal}).

The D-dimensional Riemannian space of constant negative curvature $H^D$
(hyperbolic space) can be described as a hyperboloid
\begin{equation}
- x_0^2 + x_1^2 + x_2^2 + ... + x_D^2 = - a ^{2}
\end{equation}
embedded in (D+1)-dimensional Minkowski space with metric $ds^2=-dx_0^2 + dx_1^2 + ... + dx_D^2$.
It is easy to show that the Minskowski metric becomes positive definite on the surface given by Eq.(8)
(this is the reason why we have
chosen the Minkowski metric in the form (--, +, +,..., +)). Obviously
by construction hyperbolic space has the group of
isometry SO(1, D) and is a homogeneous space because any two points on $H^D$ can be connected by some isometry
(all points of this space are equivalent).
By using the parametrization
\begin{eqnarray}
x_0 = a\cosh \sigma, \,\,\,
x_1 = a \sinh \sigma \cos \theta_1, \nonumber \\
x_2 = a \sinh \sigma \sin \theta_1 \cos \theta_2, \,\,\,
x_3 = a \sinh \sigma \sin \theta_1 \sin \theta_2 ,\,.\,.\,.\,,
\end{eqnarray}
the line element $ds^2=-dx_0^2 + dx_1^2 + ... + dx_D^2$ becomes
\begin{equation}
     ds^{2}=a^{2}(d\sigma^{2}+\sinh^{2}\sigma d\Omega_{D-1}),
\end{equation}
where $d\Omega_{D-1}$ is the metric on unit (D-1)-dimensional sphere and the curvature is equal to
\begin{equation}
     R= - \frac{D(D-1)}{a^2}.
\end{equation}
Recall that for Euclidean space the linear element in spherical coordinates is
\begin{equation}
ds^{2}=(dr^{2}+r^{2}d\Omega_{D-1}).
\end{equation}
By comparing Eq.(10) and Eq.(12), we see that the difference between flat and hyperbolic space is that the volume of
sphere in hyperbolic space grows with radius $r$ as $a^{2}\sinh^{2}\frac{r}{a}$ instead $r^{2}$ as in flat space. 
In the present work we actually consider the (D+1)-dimensional ultrastatic spacetime $R \times H^{D}$, where the components
of metric are time independent and the conditions $g_{00}=1$ and  $g_{0i} = 0$ are true
in an appropriate system of coordinates (thus, time coordinate describes evolution of fields on $H^{D}$).

\vspace{1cm}

\section{The effective reduction.}

\vspace{0.8cm}

\subsection{Heat kernels.}

\vspace{1cm}

Since the metric on $R \times H^{D}$ is time independent, the heat kernel in Eq.(6) trivially factorizes and we are left
with the problem of calculation of the heat kernel \\ $\mbox{tr} \, <x|e^{-sH}|x>$ on the hyperbolic space $H^{D}$.
As we mentioned in Introduction it was shown that $g_{c}=0$ in spaces with constant negative
curvature. The authors of these works \cite{BK, IMO} calculated heat kernel in the closed form either using
the Schwinger method for
calculation of $<x|e^{-sH}|x>$ \cite{Sch} or expressing it through the spinor Green function, which was obtained as a
solution of the corresponding differential equation in \cite{Campo}.  Heat kernel is in a certain sense
an integral characteristic. To
reveal the underlying dynamics which gives $g_{c}=0$, we need more detailed information about the system. For this we
calculate heat kernel by summing over the eigenfunctions of the Dirac operator on $H^D$ that allows us to investigate
what dynamics is responsible for $g_{c} = 0$ in spaces with negative curvature\footnote{We would like to thank
V.P. Gusynin for suggesting this approach.}.
To calculate heat kernel in the form of sum over eigenfunctions, we use the method and the results of \cite{Byts}.
To illustrate the method, we first calculate the heat kernel for scalar field $h_{scalar} =
<x|e^{-\frac{s}{a^2}A}|x>$,
where $A = -\Delta + m^2a^2$ and $-\Delta$ is
the Laplace--Beltrami operator on $H^D$ (we will
use this heat kernel when we discuss the role of Goldstone bosons), which is given by
\begin{equation}
\Delta=\frac{\partial^{2}}{\partial \sigma^{2}}+(D-1)
\coth\sigma\frac{\partial}{\partial\sigma}+(\sinh\sigma)^{-2}\Delta_{S^{D-1}}
\:,
\end{equation}
where the last term denotes the Laplace--Beltrami operator
on the unit sphere $S^{D-1}$. If the eigenfunctions of the operator (13) are known, then
we can insert their complete set in the martix element for the heat kernel. Then the heat kernel is represented
in the form of sum over eigenfunctions
\begin{equation}
h_{scalar} = <x|e^{-\frac{s}{a^2}(-\Delta + m^2a^2)}|x> = \sum_{\lambda}
e^{-\frac{s}{a^2}(\lambda + m^2a^2)}|\phi_{\lambda}(x)|^2,
\end{equation}
where $\phi_{\lambda}$ are eigenfunctions 
($-\Delta\phi_{\lambda}=\lambda\phi_{\lambda}$). It is obvious from Eq.(13) that the equation for eigenfunctions
admits the separation of variables,
therefore, we seek them in the form
$\phi=f_{\lambda}(\sigma)Y_{lm}$, where $Y_{lm}$ are the spherical
harmonics on $S^{D-1}$
\begin{eqnarray}
\Delta_{S^{D-1}}Y_{lm}=-l(l+D-2)Y_{lm} \:.\nonumber
\end{eqnarray}

Thus, the radial wave functions satisfy the ordinary differential
equation
\begin{equation} f_{\lambda}^{''}+(D-1)\coth\sigma
f_{\lambda}^{\prime}
+\left[\lambda-\frac{l(l+D-2)^{2}}{\sinh^{2}\sigma}\right]f_{\lambda}=0
\:.\nonumber\end{equation}
The only bounded solutions of Eq.(15) are
\begin{equation}
f_{\lambda}(\sigma)= C \: \Gamma\left(\frac{D}{2}\right)
\left(\frac{\sinh\sigma}2\right)^{1-D/2} P_{-1/2+ir}^{\mu}(\cosh\sigma)
\:,
\end{equation}
where $P_{\nu}^{\mu}(z)$ are the associated
Legendre functions of the first kind
\cite{Grad},
$r=(\lambda-\rho_{D}^{2})^{1/2}$ is used as a label for the continuum
spectrum, $\rho_{D} = \frac{D-1}{2}$, and $C$ is
the normalization constant. The asymptotic behaviour of $P_{\nu}^{\mu}(z)$ for $\mid z\mid \gg 1$ is
\begin{equation}
P_{\nu}^{\mu}(z)\approx
\frac{2^{\nu}\Gamma(\nu+1/2)}{\pi^{1/2}\Gamma(\nu-\mu+1)}
z^{\nu}+\frac{\Gamma(-\nu-1/2)}{2^{\nu+1}\pi^{1/2}
\Gamma(-\nu-\mu)}z^{-\nu-1} \:,\nonumber
\end{equation}
from which we
obtain the asymptotic behavior of the eigenfunctions
\begin{equation}
f_{\lambda}(\sigma)\simeq C\frac{2^D\Gamma(D/2)\Gamma(ir)}{4\pi^{1/2}\Gamma(\rho_D+ir)}
e^{-\rho_D\sigma+ir\sigma}+h.c. \:.
\end{equation}
The radial functions are bounded at infinity provided
the parameter $r$ is real, which is equivalent to the condition
$\lambda\geq\rho_{D}^{2}$. Thus, the spectrum of the Laplace--Beltrami operator has a
gap which is determined by the curvature and depends on $D$.
Since $H^D$ is a homogeneous space (i.e. all points are equivalent), the heat kernel
does not depend on $x$ and one can use any point to calculate the heat kernel. In the spherical coordinates it is very
convenient to use the origin because as follows from the
explicit solutions (Eq.(16)) only modes with {\it l} $\: = 0$ are not equal to zero at this point.
We normalized eigenfunctions so that $f_{\lambda} (0) = C$ at $x=0$. The
invariant measure defining the scalar product between eigenfunctions is
\begin{equation}
(f_{\lambda},f_{\lambda'})=\Omega_{D-1}\int_{0}^{\infty}
f_{\lambda}^{*}f_{\lambda'}(\sinh\sigma)^{D-1}d\sigma\:,
\end{equation}
where $\Omega_{D-1}$ is the volume of the (D-1)-dimensional sphere and the factor $\sinh^{D-1}\sigma$ follows
from the square root of the determinant of metric. The normalization constant is determined from
the usual condition of normalization of eigenfunctions of continuous spectrum
\begin{equation}
(\phi_{\lambda},\phi_{\lambda^{\prime}})
=\delta(\lambda-\lambda^{\prime}).
\end{equation} 
The easiest way to calculate this scalar product and determine the normalization constant for eigenfunctions
(16) is to use the
fact that the scalar product of two eigenfunctions
is expressed through the derivative of their Wronskian $W[\cdot,\cdot]$ at an arbitrary point. For us it is
most convenient to calculate the Wronskian at infinity because we know the asymptotic behavior of eigenfunctions there.
Thus, we obtain
\begin{equation}
(f_{\lambda},f_{\lambda^{\prime}})=
\frac{2^{D-1}\pi^{\frac{D}{2}}\Gamma(\frac{N}{2})}
{(r')^2 - r^2} \lim_{\sigma\rightarrow \infty}(1-\cosh^{2} \sigma)
W[P_{-\frac{1}{2} - ir}^{\mu}(\cosh \sigma),P_{-\frac{1}{2} + ir^{\prime}}^{\mu}(\cosh \sigma)]\:,
\end{equation}
where the limit is taken in the sense of distributions.
Thus, we find
\begin{equation}
|C(r)|^2=\frac{2}{(4\pi)^{D/2}\Gamma(D/2)}
\frac{\mid\Gamma(ir+l+\rho_D)\mid^{2}}{\mid \Gamma(ir)\mid^{2}} \:
\end{equation}
for the square of the normalization constant.
Since we normalized eigenfunctions as
$f_{\lambda}(0)=C$, the heat kernel is given by
\begin{equation}
h_{scalar} = \frac{1}{a^D}\int_{0}^{\infty}e^{-\frac{s}{a^2} (r^2 + m^2a^2)}|C^2(r)| dr,
\end{equation}
where we have made the change of variables $r=\sqrt{\lambda - \rho^2_D}$.
Note that $C(r)dr$ is the measure of eigenfunctions. It defines the number
of states per unit volume in the range $dr$.

We now consider the heat kernel for the Dirac operator on $R \times H^D$. Before doing it we first discuss
what we mean by chiral symmetry in spacetimes of arbitrary dimension (we consider again spacetimes whose metric
has Minkowski spacetime signature).
In Section 2, we described the chirally invariant NJL model in four-dimensional spacetime.
As well known chiral symmetry is connected with properties of representations of the Clifford
algebra (for a good description of spinors in n-dimensional spacetime see, e.g., \cite{Sohn}).
The Clifford algebra for n-dimensional spacetime of even dimension has only one
complex, irreducible representation in the $2^{n/2}$-dimensional
spinor space. These spinors are reducible with respect to the even
subalgebra (generated by products of an even number of Dirac matrices)
and split in a pair of
$2^{n/2-1}$-component irreducible Weyl spinors ($\gamma_{n+1} = \gamma_0 ... \gamma_{n-1}$ is
an analog of the $\gamma_5$ matrix in n-dimensional spacetime and $\frac{1 \pm \gamma_{n+1}}{2}$ are the corresponding
chiral projectors). In odd-dimensional spacetimes, there are two different representations of the Clifford algebra (they
differ by the sign of the $\gamma$-matrices) and chiral symmetry is not defined because $\gamma_{n+1}$ is proportional
to the unity. In order to define chiral symmetry in odd-dimensional spacetimes, it is the usual practice to assume that
fermion fields are in a reducible representation of the Clifford algebra so that we can define an analog of chiral
symmetry (for an explicit example in
(2 + 1)-dimensional spacetime see, e.g., \cite{App}). In what follows
we understand chiral symmetry in odd-dimensional spacetimes in this sense.

In the scalar case it is easy to
factorize the
part of heat kernel, which contains time derivatives. It is a little bit more elaborated for spinors because
the time derivative is multiplied by the $\gamma_0$-matrix.
By using (see Eq.(5))
\begin{eqnarray}
(i\gamma^{\mu}\nabla_{\mu}+m)(i\gamma^{\nu}\nabla_{\nu}-m) = (i\nabla_0 + i\vec{\gamma}\gamma_0\vec{\nabla}
+m\gamma_0)\gamma_0\gamma_0(i\nabla_0 + i\gamma_0\vec{\gamma}\vec{\nabla}-m\gamma_0) = \nonumber \\
(i\nabla_0 - i\vec{\alpha}\vec{\nabla}+m\gamma_0)(i\nabla_0 + i\vec{\alpha}\vec{\nabla}-m\gamma_0) =
(i\nabla_0)^2 - (-i\vec{\alpha}\vec{\nabla}+m\gamma_0)^2,
\end{eqnarray}
where $\vec{\alpha} = \gamma_0 \vec{\gamma}$, we get rid of the $\gamma_0$-matrix.
It is no wonder why such an operator
for the heat
kernel for spatial coordinates appears after the separation
of time derivatives. Indeed, the Dirac equation can be written in
the form of a Schr\"{o}dinger equation
\begin{eqnarray}
i\frac{\partial \psi}{\partial t} = H \psi \nonumber
\end{eqnarray}
with the Hamiltonian $H = -i\vec{\alpha}\vec{\nabla} + \beta m$,
where $\vec{\alpha}=\gamma_0\vec{\gamma}$ and $\beta = \gamma_0$.
Therefore, we immediately recognize our operator $(-i\vec{\alpha}\vec{\nabla}+m\gamma_0)^2$ as the square
of the Hamiltonian,
which is obviously a positive definite operator. Thus, the gap equation (see Eq.(7)) on $R \times H^D$ is
\begin{equation}
1\,\, = \,\,G\int_{\frac{1}{\Lambda^2}}^{\infty} ds \,\mbox{tr}
<t,x|e^{-s(-(\nabla_0)^2 + (-i\vec{\alpha}\vec{\nabla}+m\gamma_0)^2)}|t,x>,
\end{equation}
where we again performed the Wick rotation.
The contribution of time derivatives to the heat kernel is easy to be found.
Therefore, we are left with the problem of calculation of heat
kernel on $H^D$ for the operator $(-i\vec{\alpha}\vec{\nabla}+m\gamma_0)^2$.

To calculate the heat kernel for the operator $(-i\vec{\alpha}\vec{\nabla}+m\gamma_0)^2$ we use as in the scalar case
expansion in eigenfunctions.
The equation for eigenfunctions is
\begin{eqnarray}
(-i\vec{\alpha}\vec{\nabla} + m\gamma_0)\psi_{\lambda}=\lambda\psi_{\lambda} \nonumber\:.
\end{eqnarray}
The covariant derivative of a spinor field on $H^{D}$ can be
decomposed in a radial part plus the covariant derivative along the
unit $S^{(D-1)}$-sphere. Furthermore,
making the decomposition of $2^{\frac{D+1}{2}}$-dimensional representation in a Dirac-like
representation of $\gamma$-matrices, the equation for eigenfunctions takes the form of a
coupled system (for more details see \cite{Byts})
\begin{equation}
i\gamma_{1}\left(\partial_{\sigma}+\rho_D
\coth\sigma\right)\psi_{1}+\frac{1}{\sinh\sigma}i\not\!\nabla_s\psi_{1}
=-a(\lambda+m)\psi_{2} \:,
\end{equation}
\begin{equation}
i\gamma_{1}\left(\partial_{\sigma}+\rho_D
\coth\sigma\right)\psi_{2}+\frac{1}{\sinh\sigma}i\not\!\nabla_s\psi_{2}
=-a(\lambda-m)\psi_{1} \:,
\end{equation} where
$\psi_{1,2}$ are the $2^{\frac{D+1}{2}-1}$-components Weyl spinors and $i\not\!\nabla_s$ is the
Dirac operator on $S^{D-1}$. The spinors $\psi_{1,2}$ transform
irreducibly under $SO(D)$ so that we can put
$\psi_{1,2}=f_{1,2}(\sigma)\chi_{1,2}$, where $\chi_{1,2}$ are spinors
on $S^{D-1}$.
The eigenvalues of the Dirac operator on $S^{D-1}$ are known to be
$\kappa=\pm(l+\rho_D)$, $l=0,1,2,...$ \cite{Cand}.
The solutions of Eqs.(26) and (27) are given in terms
of hypergeometric functions as follows:
\begin{eqnarray}
f_{1}^{+}(\sigma)&=&A\frac{ia(\lambda+m)}{l+N/2}\left(\frac{\lambda-m}{4 \lambda}\right)^{\frac{1}{2}}
(1+z)^{\frac{l}{2}}(z-1)^{\frac{l+1}{2}}F\left(\alpha,\alpha^*;
l+\rho_N+\frac{3}{2};\frac{1-z}{2}\right)\:,\nonumber\\
f_{2}^{+}(\sigma)&=&A\left(\frac{\lambda-m}{4 \lambda}\right)^{\frac{1}{2}}(1+z)^{\frac{l+1}{2}}(z-1)^{\frac{l}{2}}
F\left(\alpha,\alpha^*;l+\rho_N+
\frac{1}{2};\frac{1-z}{2}\right)\:,\nonumber\\
f^{-}_{1}(\sigma)&=&A\left(\frac{\lambda+m}{4 \lambda}\right)^{\frac{1}{2}}(1+z)^{\frac{l+1}{2}}(z-1)^{\frac{l}{2}}
F\left(\alpha,\alpha^*;l+\rho_N+
\frac{1}{2};\frac{1-z}{2}\right)\:, \\
f^{-}_{2}(\sigma)&=&A\frac{ia(\lambda-m)}{l+N/2}\left(\frac{\lambda+m}{4 \lambda}\right)^{\frac{1}{2}}
(1+z)^{\frac{l}{2}}(z-1)^{\frac{l+1}{2}}F\left(\alpha,\alpha^*;
l+\rho_N+\frac{3}{2};\frac{1-z}{2}\right) \:, \nonumber
\end{eqnarray}
where $f_{1,2}^{\pm}(r)$ are the
solutions with $\kappa =\pm(l+\rho_D)$, $z=\cosh \sigma$, $\alpha=l+D/2+ir$,
$r=a\sqrt{\lambda^2 - m^2}$, and $A$ is the normalization constant.
As in the scalar case the normalization constant $A$ is determined from the usual $\delta$-function
condition of normalization of eigenfunctions of continuous spectrum
\begin{equation}
|A(r)|^2=\frac{\Gamma(\frac{D}{2})}{\pi^{\frac{D}{2}+1}2^{D+1+2l}}
\frac{|\Gamma(D/2+l+ir)|^2|\Gamma(ir)|^2}{|\Gamma(2ir)|^2}\:.
\end{equation}
Note that the
solutions remain bounded at infinity if $r$ is real. Hence, the
spectrum of the Dirac operator on $H^D$ is $|\lambda|\geq m$.
Thus, unlike the scalar case, there is no gap for fermions on $H^D$ (this fact is very important
for what follows). Nevertheless,
the solutions are exponentially vanishing at infinity.
Having determined the normalization constant, we immediately get the heat kernel for spinors on $H^D$
(again as in the scalar case only modes with $l\,=\,0$ are not equal to zero at the origin) 
\begin{equation}
h_{H^D} \, = \,\frac{2^{[\frac{D+1}{2}]}}{a^D}\int_0^\infty
e^{-\frac{s}{a^2}(r^2+m^2a^2)}|A(r)|^2dr\,,
\end{equation}
where $[\frac{D+1}{2}]$ denotes the
integer part of $\frac{D+1}{2}$, which results from the trace over the spinor indices.
(Note that this heat kernel calculated by 'brute force' through summation over eigenfunctions coincides
with the heat kernel calculated in \cite{Campo}, which is expressed through the spinor Green function obtained as a
solution of the corresponding differential equation).

\vspace{1cm}

\subsection{Analysis.}

\vspace{1cm}

To interprete the heat kernel obtained, we remind the results of the corresponding calculations in flat spacetime.
The gap equation in flat spacetime is
\begin{equation}
1\,\, = \,\,G\int_{\frac{1}{\Lambda^2}}^{\infty} ds \,\, h_{flat},
\end{equation}
where $h_{flat} \, = \, tr<x|e^{-s(-(\gamma_E^{\mu}\nabla_{\mu})^2 + m^2)}|x>$.
The eigenfunctions of the Dirac operator in flat spacetime are just plane waves, therefore, the corresponding
heat kernel is
\begin{equation}
h_{flat} \, = \,2^{\frac{n}{2}}\int \frac{d^n k}{(2\pi)^n}
e^{-s(k^2+m^2)}.
\end{equation}
By integrating over angular variables,
we obtain
\begin{equation}
h_{flat}\, = \, 2^{\frac{n}{2}}\int_0^{\infty}
\frac{2 \, dk k^{n-1}}{(4 \pi)^{\frac{n}{2}} \Gamma(\frac{n}{2})} e^{-s(k^2+m^2)} =
\frac{2^{\frac{n}{2}}e^{-s m^2}}{(4 \pi s)^{\frac{n}{2}}}
\end{equation}
Thus, we see that the function $k^{n-1}$ defines
a measure in space of eigenfunctions and for $m = 0$ determines the asymptotic behavior of the heat kernel at large s.
Obviously, every new dimension gives an additional factor
$s^{-\frac{1}{2}}$ to the asymptotic behavior of
the heat kernel. Furthermore, we see from the gap equation (31) that in two-dimensional spacetime $G \to 0$ if $m \to 0$
because the integral over $s$ is divergent on the upper limit in this case.  Thus, the critical value
of coupling constant is zero.
We now return to the heat kernel on $R \times H^D$. It is
\begin{equation}
h_{R \times H^D} \, = \, \frac{2^{[\frac{D+1}{2}]}}{a^D}\int_0^\infty
\frac{e^{-\frac{s}{a^2}(r^2+m^2a^2)}}{(4 \pi s)^{\frac{1}{2}}}|A(r)|^2dr\,,
\end{equation}
where the factor $(4 \pi s)^{\frac{1}{2}}$ is the contribution to the heat kernel from time coordinate
and the rest is the heat kernel on $H^D$ (see Eq.(30)).
Let us explicitly calculate the measure $|A(r)|^2$, which is given by Eq.(29) with $l\, = \, 0$.
By using the formulas \cite{Grad}
\begin{eqnarray}
|\Gamma(iy)|^2 = \frac{\pi}{y\sinh(\pi y)}, \nonumber \\
|\Gamma(\frac{1}{2} + iy)|^2 = \frac{\pi}{\cosh(\pi y)},
\end{eqnarray}
we get
\begin{equation}
|A(r)|^2 = \frac{r \coth(\pi r) \: \prod_{j=1}^{\frac{D-2}{2}} \: (r^2 + j^2)}
{\pi^{\frac{D}{2}}2^{D-1} \Gamma(\frac{D}{2})}
\end{equation}
for even D and
\begin{equation}
|A(r)|^2 = \frac{\prod_{j=\frac{1}{2}}^{\frac{D-2}{2}} \: (r^2 + j^2)}
{\pi^{\frac{D}{2}}2^{D-1} \Gamma(\frac{D}{2})}
\end{equation}
for odd D.

The asymptotic behavior of the heat kernel for large s in the case of critical coupling
constant ($m = 0$) is determined by the behavior of the integrand at small $r$.
As follows from Eqs. (36) and (37), for small $r$, the measure $|A(r)|^2$ tends to a constant for any D.
Consequently, we obtain that the leading term of the heat kernel (34) is $h_{R \times H^D} \sim \frac{1}{s}$.
Thus, the fermion dynamics on $R \times H^D$ in the infrared region corresponds to the dynamics of
(1 + 1)-dimensional theory.
(Note that in the opposite limit of small s (large energies that corresponds to large $r$)
the measure $|A(r)|^2$ tends to
$\frac{2r^{D-1}}{(4 \pi)^{\frac{D}{2}} \Gamma(\frac{D}{2})}$.
Therefore, the leading term of the heat kernel on $R \times H^D$ at $s \to 0$ is
$\frac{1}{(4 \pi s)^{\frac{D+1}{2}}}$ that corresponds to the behavior of (D+1)-dimensional theory
as expected). Consequently, we can say that the effective reduction of the dimension of spacetime
$1 + D \to 1 + 1$ takes place in the infrared region for fermion fields
for any 1 + D. This explains why $g_c = 0$ in spaces with constant negative curvature (it immediately follows
from the gap equation if $h \sim \frac{1}{s}$ for large $s$). For
completeness we present the corresponding results of the effective reduction for the case of the NJL model in
four-dimensional spacetime in external magnetic field \cite{GMSh1, GMSh2}.
The heat kernel for the Dirac operator in constant magnetic field is
\begin{eqnarray}
h_{magnetic}\, = \, \frac{e^{-s m^2} eB \cot (eBs)}{16 \pi^2 s}\,,
\end{eqnarray}
where $e$ is the charge of the electron and $B$ is magnetic field.
The heat kernel (38) evidently also corresponds to
(1 + 1)-dimensional theory in the infrared because $\coth(eBs)$ tends to 1 for large s.

\vspace{1cm}

\subsection{Nambu--Goldstone bosons.}

\vspace{1cm}

In the preceding subsection we found that the dynamics of fermions in the infrared is effectively (1 + 1)-dimensional.
Potentially, it may present a problem for dynamical symmetry breaking of a continuous symmetry because,
according to the Coleman--Mermin--Wagner theorem \cite{Cole}, spontaneous symmetry breaking of a continuous symmetry
is not possible in 1 + 1 due to strong infrared divergences connected with massless Nambu--Goldstone bosons
(the existence of
this potential problem in theories with the effective reduction of dimension of spacetime was indicated in
\cite{GMSh1, GMSh2}).
For example, in the NJL model the following diagram of the next-to-leading (in $\frac{1}{N}$) correction to vacuum energy
is infrared divergent in 1 + 1:

\begin{figure}[htb]
\begin{center}
\begin{picture}(200,100)(0,0)
\CArc(100,50)(50,0,360)
\DashLine(50,50)(150, 50){5}
\end{picture}
\end{center}
\end{figure}

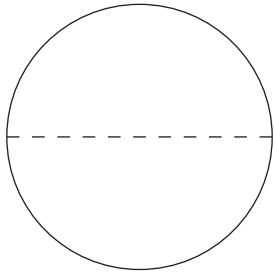
Fig. 1.  The next-to-leading order correction to vacuum energy in the NJL model. The fermion propagators
are denoted by solid lines. A dashed line denotes the propagators of $\sigma$ and $\pi$ in the leading order in $1/N$.

\vspace{1cm}

If the effective reduction of dimension of spacetime in the infrared region took place for scalars, then we
would have a problem connected with infrared divergent radiative corrections due to massless Nambu--Goldstone bosons.
For the case of the
effective reduction in external magnetic
field, Gusynin, Miransky, and Shovkovy \cite{GMSh1, GMSh2} presented an elegant solution of this potential problem.
They indicated
that since in the case of chiral symmetry breaking the condensate $<0|\bar{\psi}\psi|0>$ is neutral and the
Nambu--Goldsone bosons are
neutral particles, the effective dimensional reduction (which for fermions reflects the fact that the motion
of charged particles is restricted in the directions perpendicular to the magnetic field) does not affect the dynamics
of the center of mass of neutral excitations.
Therefore, as they showed by explicit calculations the propagators of Nambu--Goldstone bosons have
(3 + 1)-dimensional form in the infrared region. Evidently such a solution
cannot be used in the case of gravitational field because gravity is universal and
all particles including Nambu--Goldstone bosons directly interact
with gravitational field. Therefore, we should seek another solution. For this end we consider
the propagator of massless
scalar field. This propagator can be expressed through the nondiagonal heat kernel of the Laplace--Beltrami operator.
Time dependence is trivially factorized and we are left with the problem of
calculating heat kernel on $H^D$. In Subsection 3.1 we have calculated the diagonal
heat kernel $h_{scalar} = <x|e^{-\frac{s}{a^2}(-\Delta)}|x>$ (see Eq.(23)).
The nondiagonal heat kernel was calculated in
\cite{Cam}
\begin{equation}
<x|e^{-\frac{s}{a^2}(-\Delta)}|y> = \frac{1}{a^D} \int_0^{\infty} \phi_r (\tau) |C(r)|^2
e^{-\frac{s}{a^2}(r^2 + \rho_D^2)} dr,
\end{equation}
where $\phi_r (\tau) = F(ir + \rho_D, -ir + \rho_D, \frac{D}{2}; - \sinh^2 \frac{\tau}{2a})$ ($\tau$ is the geodesic
distance between points $x$ and $y$) and
$|C(r)|^2$ is given by Eq.(22).
Thus, we obtain the propagator for scalar massless particles
\begin{equation}
G(t-t^{\prime}, \tau) = \frac{1}{a^D}\int^{\infty}_{1/\Lambda^2}
ds \frac{e^{-\frac{t-t^{\prime}}{4s}}}{(4\pi s)^{\frac{1}{2}}}
\int_0^{\infty} \phi_r (\tau) |C(r)|^2
e^{-\frac{s}{a^2}(r^2 + \rho_D^2)} dr\,,
\end{equation}
where we have performed the Wick rotation in time coordinate.

Obviously, since there is a gap in the spectrum of the Laplace--Beltrami operator, there are not any problems with infrared
behavior of massless scalar particles.
Indeed, in the proper time method infrared divergences are connected with the
divergence of the integral over $s$ on the upper
limit of integration. Since there is a gap in the spectrum, the integrand has the factor
$e^{-\frac{s}{a^2}\rho_D^2}$. Therefore,
infrared divergences are absent.
Thus, we conclude that the effective reduction of dimension of spacetime in
the infrared region for fermion fields does not contradict the Mermin--Wagner--Coleman theorem.
Note that our calculations show that there are not gapless bosonic modes. Consequently, there are not gapless
Nambu--Goldstone bosons in this model. However, this does not contradict the Goldstone theorem: this theorem
has been proved only for
Minkowski space. The problem of a possible extending the theorem to the case of curved spacetime will be considered
elsewhere.

\vspace{1cm}

\section{Conclusion.}

\vspace{1cm}

In the present paper we studied chiral symmetry breaking in the NJL model
in spaces with constant negative curvature. We showed that
zero value of critical coupling constant $g_c = 0$ is connected with the effective reduction of dimension of spacetime 
$1 + D \to 1 + 1$ for fermions in the infrared region.
Note that this effective reduction has a universal
character
in the sense that the initial theory reduces in the infrared region to two-dimensional one in the fermion sector
for any dimension $1 + D$. In this respect
this is similar to the effective reduction in the presence of the Fermi
surface when the net fermion charge is not equal to zero.

By analysing the scalar propagator,
we showed that such an effective reduction is absent in the scalar sector, therefore, the
effective reduction of the dimension of spacetime for fermions and symmetry breaking are consistent
and there is not a contradiction with
the Coleman--Mermin--Wagner theorem, which states that spontaneous symmetry breaking is not possible
in 1 + 1.

Finally let us mention that the hyperbolic space $H^D$ is an Euclidean analog of anti-de Sitter space
(the Wick rotation of AdS gives
$H^D$). Recently the dynamics of quantum fields on AdS has received a lot of attention in view of the conjectured CFT/AdS
correspondence \cite{Mal}. Therefore, it is a natural problem to study what the dynamics we discuss here means
in the context of this correspondence. The results of this study will be published elsewhere.

The author thanks V.P. Gusynin for the suggestion to use expansion of the heat kernel in eigenfunctions of the
Dirac operator and for the
critique of an earlier version of this paper. The author thanks V.A. Miransky for useful remarks and suggestions and
also acknowledges helpful discussions with
A.S. Belyaev, M. Nowakowski, L.C.B. Crispino, I.L. Shapiro, and Yu.V. Shtanov.
I am grateful to I.L. Shapiro for bringing my attention to \cite{Byts}.
This work was supported in part by FAPESP grant No. 98/06452-9.


\begin{thebibliography}{99}

\bibitem{book} V. A. Miransky, {\it Dynamical Symmetry Breaking in
Quantum Field Theories},
(World Scientific, Singapore, 1993).
\bibitem{BCS} J. Bardeen, L.N. Cooper, and J.R. Schrieffer, Phys.Rev. {\bf 108} (1957) 1175.
\bibitem{Pol} R. Shankar, Rev.Mod.Phys. {\bf 66} (1994) 129; \\
J. Polchinski, "Effective Field Theory and the Fermi Surface," in Proceedings of the 1992 TASI, eds. J. Harvey and J.
Polchinski (World Scientific, Singapore 1993), hep-th/9210046.
\bibitem{Alf} M. Alford, K. Rajagopal, and F. Wilczek, Phys.Lett. {\bf B422} (1998) 247.
\bibitem{Rap} R. Rapp, T. Sch\"{a}fer, E.V. Shuryak, and M. Velkovsky, Phys.Rev.Lett. {\bf 81} (1998) 53.
\bibitem{GMSh1} V.P. Gusynin, V.A. Miransky, and I.A. Shovkovy, Phys.Rev.Lett. {\bf 73} (1994) 3499; \\
V.P. Gusynin, V.A. Miransky, and I.A. Shovkovy, Phys.Lett. {\bf B349} (1995) 477; \\
V.P. Gusynin, V.A. Miransky, and I.A. Shovkovy, Phys.Rev. {\bf D52} (1995) 4718.
\bibitem{GMSh2} V.P. Gusynin, V.A. Miransky, and I.A. Shovkovy, Phys.Rev. {\bf D52} (1995) 4747; \\
V.P. Gusynin, V.A. Miransky, and I.A. Shovkovy, Nucl.Phys. {\bf B462} (1996) 249.
\bibitem{NJL} Y. Nambu and G. Jona-Lasinio, Phys.Rev. {\bf 122} (1961) 345.
\bibitem{Kri} I.V. Krive and S.A. Naftulin, Phys.Rev. {\bf D46} (1992) 2737; \\
K.G. Klimenko, Theor.Math.Phys. {\bf 89} (1992) 1161.
\bibitem{Kle} S.P. Klevansky, Rev.Mod.Phys. {\bf 64} (1992) 1.
\bibitem{BK} I.L. Buchbinder and E.N. Kirilova, Int.J.Mod.Phys. {\bf A4} (1989) 143; \\
C.T. Hill and D.S. Salopek, Ann.Phys. (NY) {\bf 213} (1992) 21; \\
T. Muta and S.D. Odintsov, Mod.Phys.Lett. {\bf A6} (1991) 3641.
\bibitem{IMO} T. Inagaki, T. Muta, and S.D. Odintsov, Mod.Phys.Lett. {\bf A8} (1993) 2117; \\
I. Sachs and A. Wipf, Phys.Lett. {\bf B326} (1994) 105; \\
E. Elizalde, S.D. Odintsov, and Yu.I. Shil'nov, Mod.Phys.Lett. {\bf A9} (1994) 913; \\
S. Kanemura and H.-Y. Sato, Mod.Phys.Lett. {\bf A11} (1996) 785; \\
B. Geyer, L.N. Granda, and S.D. Odintsov, Mod.Phys.Lett. {\bf A11} (1996) 2053; \\
T. Inagaki, Int.J.Mod.Phys. {\bf A11} (1996) 4561; \\
E. Elizalde, S. Leseduarte, S.D. Odintsov, and Yu.I. Shil'nov, Phys.Rev. {\bf D53} (1996) 1917; \\
D.M. Gitman, S.D. Odintsov, and Yu.I. Shil'nov, Phys.Rev. {\bf D54} (1996) 2968; \\
T. Inagaki and K.-I. Ishikawa, Phys.Rev. {\bf D56} (1997) 5097; \\
G. Miele and P. Vitale, Nucl.Phys. {\bf B494} (1997) 365.
\bibitem{Ina} T. Inagaki, T. Muta, and S.D. Odintsov, Prog.Theor.Phys.Suppl. {\bf 127} \\
(1997) 93.
\bibitem{Wil} C.G. Callan and F. Wilczek, Nucl.Phys. {\bf B340} (1990) 366.
\bibitem{Hub} J. Hubbard, Phys.Rev.Lett. {\bf 3} (1959) 77.
\bibitem{Stra} R.L. Stratonovich, Sov.Phys.-Dokl. {\bf 2} (1958) 416.
\bibitem{Sch} J. Schwinger, Phys. Rev. {\bf 82} (1951) 664.
\bibitem{Bal} N.L. Balazs and A. Voros, Phys.Rep. {\bf C143} (1986) 109.
\bibitem{Campo} R. Camporesi, Commun.Math.Phys. {\bf 148} (1992) 283.
\bibitem{Byts} A.A. Bytsenko, G. Cognola, L. Vanzo, and S. Zerbini, Phys.Rep. {\bf C266} \\ (1996) 1.
\bibitem{Grad} I.S. Gradsteyn and I.M. Ryzhik, {\it Tables of Integrals, Series, and Products}, \\
(Academic Press, New York, 1965).
\bibitem{Sohn} M.F. Sohnius, Phys.Rep. {\bf C128} (1985) 39, Appendix A.
\bibitem{App} T. Appelquist, M.J. Bowick, D. Karabali, and L.C.R. Wijewardhana, Phys.Rev. {\bf D33} (1986) 3774.
\bibitem{Cand} P. Candelas and S. Weinberg, Nucl.Phys. {\bf B237} (1984) 397.
\bibitem{Cole} N. D. Mermin and H. Wagner, Phys. Rev. Lett. {\bf 17}
(1966) 1133;\\
 S. Coleman, Commun. Math. Phys. {\bf 31} (1973) 259.
\bibitem{Cam} R. Camporesi, Phys.Rev. {\bf D43} (1991) 3958.
\bibitem{Mal} J. Maldacena, Adv.Theor.Math.Phys. {\bf 2} (1998) 231, hep-th/9711200;\\
S.S. Gubser, I.R. Klebanov, and A.M. Polyakov, Phys.Lett. {\bf B428} (1998) 253, hep-th/9802109; \\
E. Witten, Adv.Theor.Math.Phys. {\bf 2} (1998) 253, hep-th/9802150. 

\end{thebibliography}
\end{document}